\documentclass[a4paper]{VisionStyle}
\usepackage{epsfig}

\begin{document}

\title{Ultraviolet images of Seyfert galaxies from the Optical Monitor on XMM-Newton}

\author{A.J.\,Blustin\inst{1} \and G.\,Branduardi-Raymont\inst{1} \and 
  A.\,Breeveld\inst{1} \and A.\,Brinkman\inst{2} \and S.\,Kahn\inst{3} } 

\institute{MSSL, University College London,
             Holmbury St. Mary, Dorking, Surrey RH5 6NT, England
\and 
  SRON, Sorbonnelaan 2, 3584 CA Utrecht, Netherlands
\and
  Columbia University, New York, NY 10027, USA}

\maketitle 

\begin{abstract}

The Optical Monitor telescope (\cite{ablustin-C2_27:mas01}) on XMM-Newton provides an exciting multi-wavelength dimension to observations of Active Galactic Nuclei. Here we present ultraviolet images, taken with the OM UVW2 filter (140-270 nm), of various Seyfert galaxies, some of which have never been observed in this waveband before. The images show UV emission from both the active nucleus and the host galaxy. The distribution of UV emission in the galaxy shows where star formation is occurring, thus giving us clues as to the evolution of the host galaxy and perhaps its relationship to the Seyfert Nucleus.

\end{abstract}

\section{NGC 7314 and NGC 7313}
\label{ablustin-C2_27_sec:7314_7313}

\begin{figure*}
  \begin{center}
    \epsfig{file=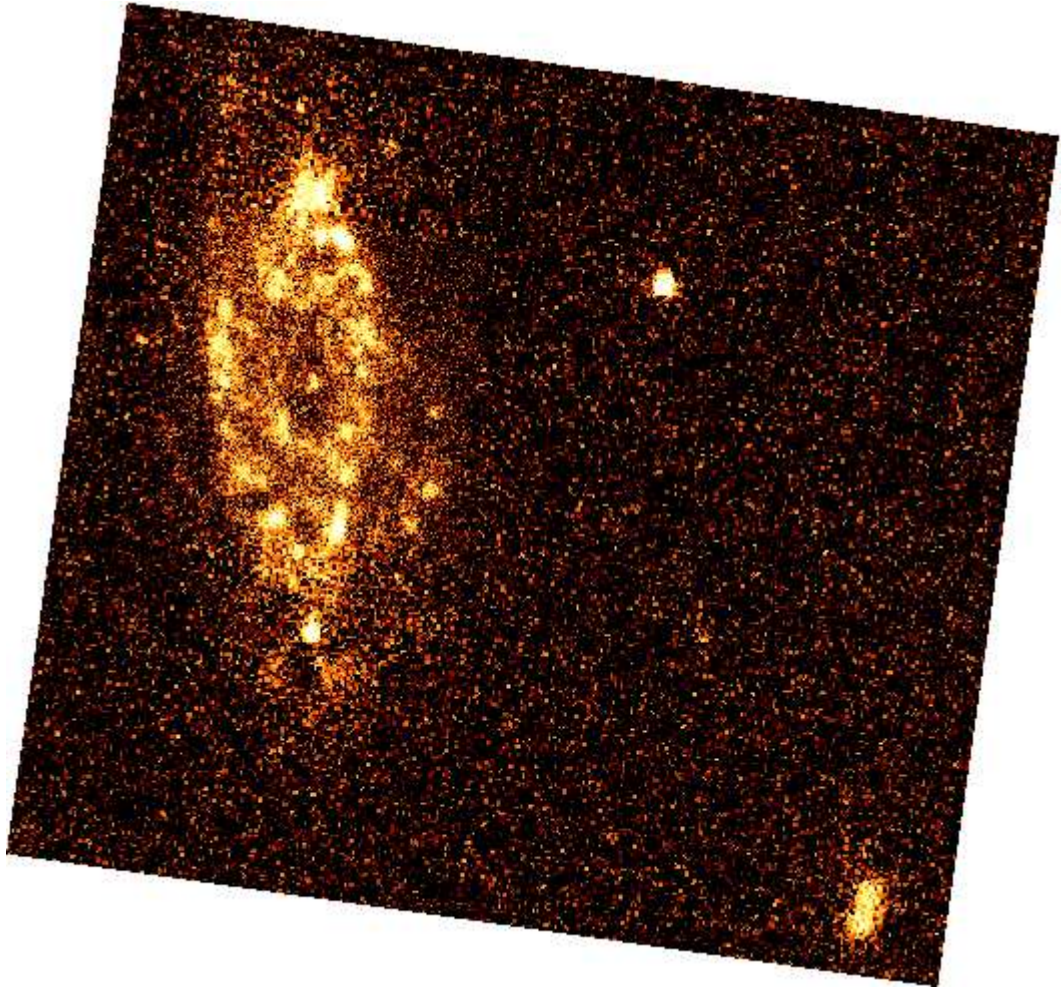, width=14cm}
  \end{center}
\caption{UV image of \object{NGC~7314} (top left) and \object{NGC~7313} (bottom right); image rotated so that North is upwards and East is to the left (all the images in this paper are oriented the same way), image size approx. 6\arcmin x 5\arcmin. The angular resolution of the image is 1\arcsec, except for the central 2\arcmin \, of \object{NGC~7314} which is at 0.5\arcsec \, resolution }  
\label{ablustin-C2_27_7314_7313:fig1}
\end{figure*}

The UV image of \object{NGC~7314} (Figure~\ref{ablustin-C2_27_7314_7313:fig1}; top left), a Seyfert 1.9 galaxy at z=0.00474, shows that its active nucleus is far less bright in UV than the spiral arms, where the UV emission mostly coincides with the locations of HII regions (and optical emission) in the galaxy. The Seyfert nucleus sits in a mostly UV-dark region at the centre of the galaxy. Overall, the spiral arms are observed to extend about 1\arcmin \, horizontally (as seen here) and 2.7\arcmin \, vertically. We also see in this image the spiral galaxy \object{NGC~7313} (z=0.01915), bottom right, 4.3\arcmin \, away from the nucleus of \object{NGC~7314}. The bright object at the top right is a Galactic star.

\section{NGC 3783}
\label{ablustin-C2_27_sec:3783}

\begin{figure*}
  \begin{center}
    \epsfig{file=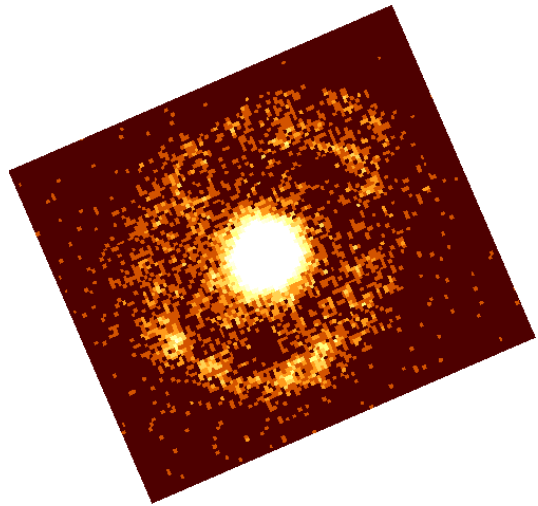, width=7cm}
  \end{center}
\caption{UV image of \object{NGC~3783}; image size approx. 50\arcsec x 40\arcsec, angular resolution 0.5\arcsec}  
\label{ablustin-C2_27_3783:fig2}
\end{figure*}

Figure~\ref{ablustin-C2_27_3783:fig2}, the first ever image to be taken of \object{NGC~3783} (z=0.00973) in this waveband, shows UV emission from the Seyfert 1 nucleus and from the spiral arms of the galaxy which are 30\arcsec \, in diameter. There is clearly discernable structure in the UV, presumably tracing the location of star formation. Although the spiral arms are visible, the bar of the galaxy (seen in optical images running approximately north to south) is not strongly present, indicating that star formation is much less important in the bar than it is in the arms, or that it is obscured by dust.

\section{NGC 7469 and IC 5283}
\label{ablustin-C2_27_sec:7469_5283}

\begin{figure*}
  \begin{center}
    \epsfig{file=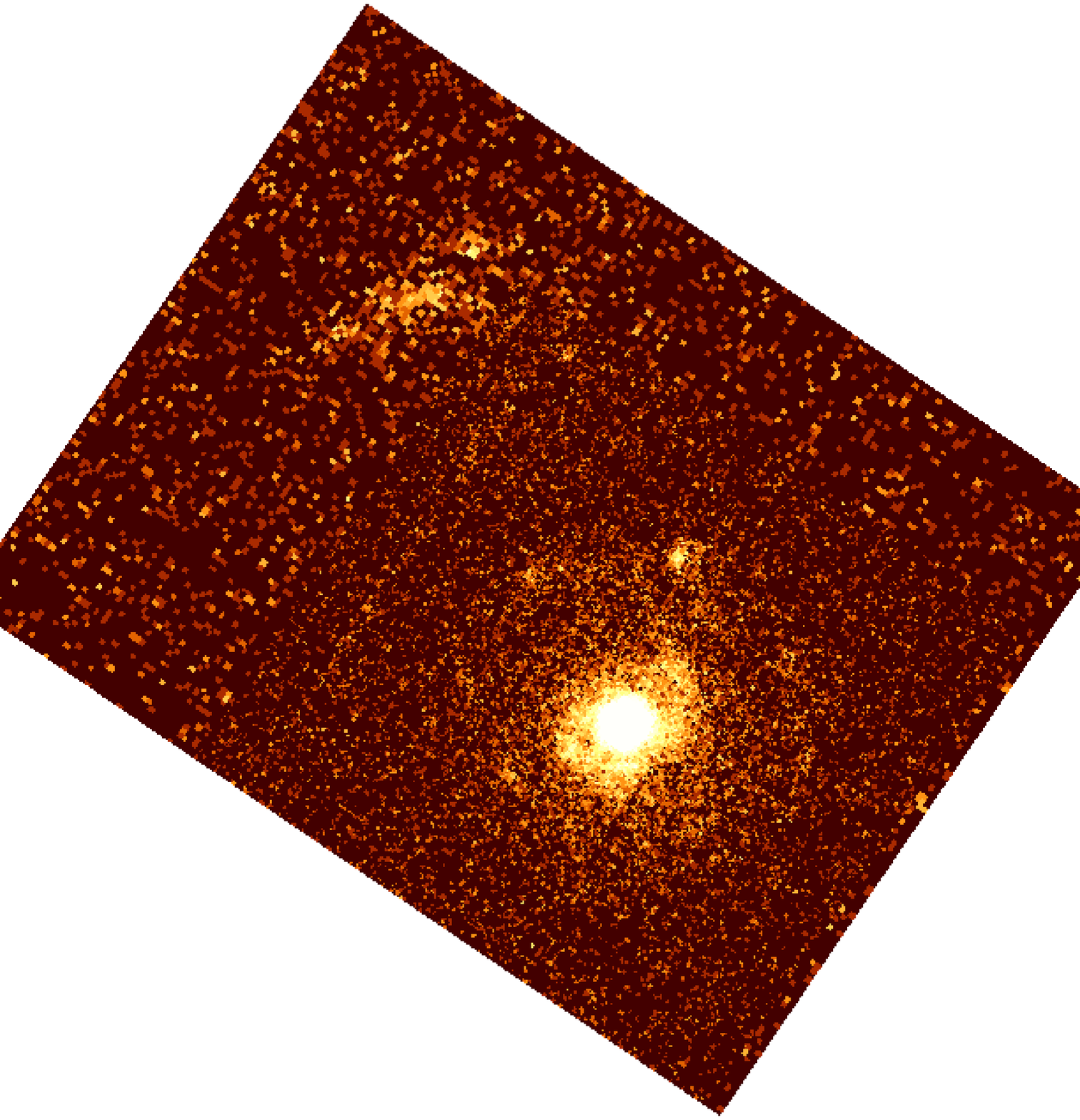, width=12cm}
  \end{center}
\caption{UV image of \object{NGC~7469} (bottom right) and \object{IC~5283} (top left); image size approx. 140\arcsec x 110\arcsec. Note that \object{NGC~7469} is within a higher resolution imaging window (with 0.5\arcsec \, imaging pixels), whereas \object{IC~5283} is imaged at 1\arcsec \, angular resolution }   
\label{ablustin-C2_27_7469_5283:fig3}
\end{figure*}

This image (Figure~\ref{ablustin-C2_27_7469_5283:fig3}) shows UV emission from the active nucleus of \object{NGC~7469} (z=0.016317; bottom right), and also from its inner spiral arms. These extend up to about 15\arcsec \, from the nucleus, and fainter arms can be observed further out at about 30\arcsec. We also see (top left) UV emission from the interacting companion galaxy \object{IC~5283} (z=0.016024), 1.3\arcmin \, away. Some of the non-AGN emission from \object{NGC~7469} is probably due to reflection of - or ionisation by - the radiation from the nucleus itself, and some of it undoubtedly traces the location of star formation activity; this galaxy is well-known as both the host of a Seyfert 1 nucleus and a starburst. The knots in the outer spiral arms seem to coincide with HII regions.

\section{Markarian 766 and MCG-6-30-15}
\label{ablustin-C2_27_sec:mrk_mcg}

\begin{figure}
  \begin{center}
    \epsfig{file=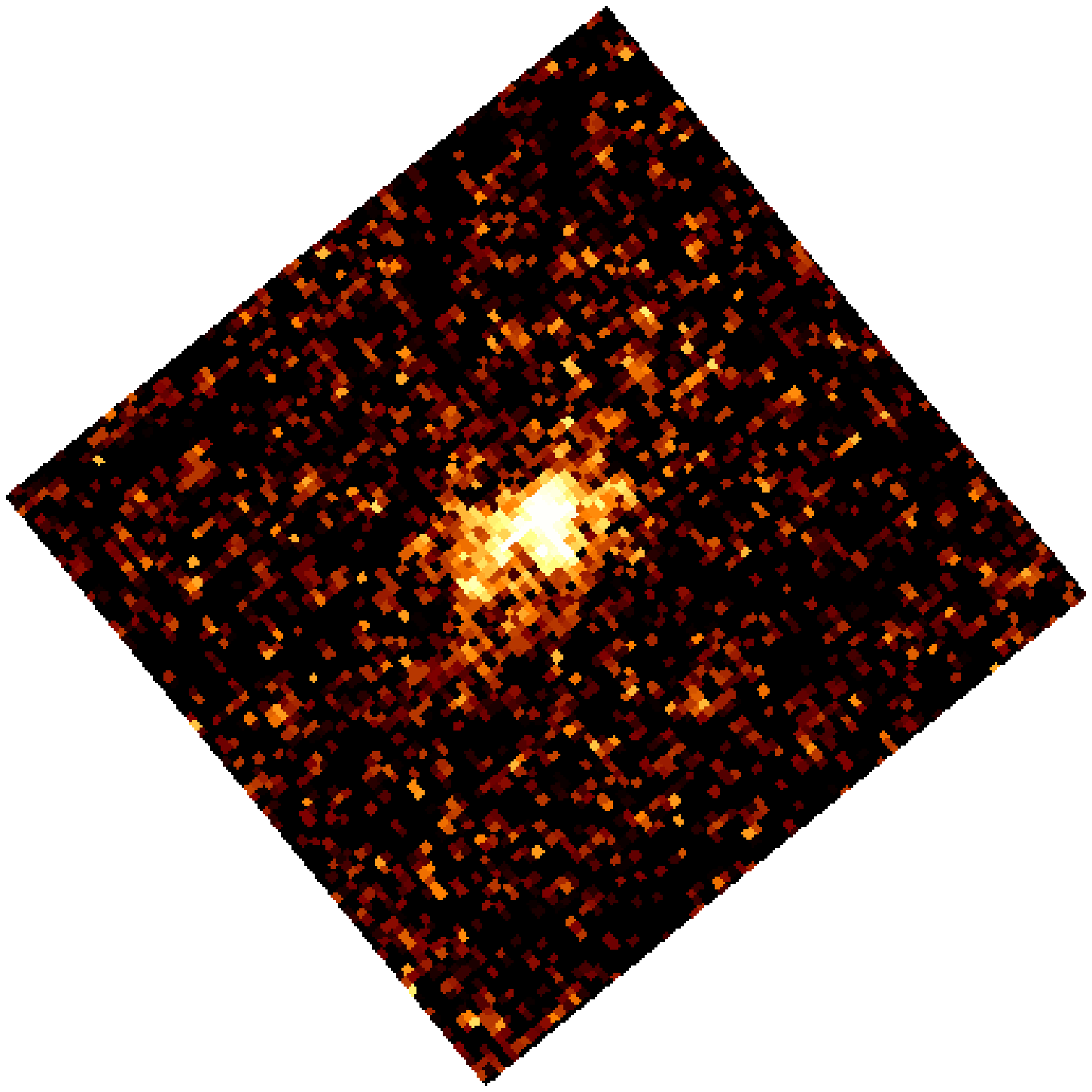, width=7cm}
  \end{center}
\caption{UV image of \object{Markarian~766}; image approx. 38\arcsec \, square, angular resolution 0.5\arcsec }  
\label{ablustin-C2_27_mrk:fig4}
\end{figure}

\begin{figure}
  \begin{center}
    \epsfig{file=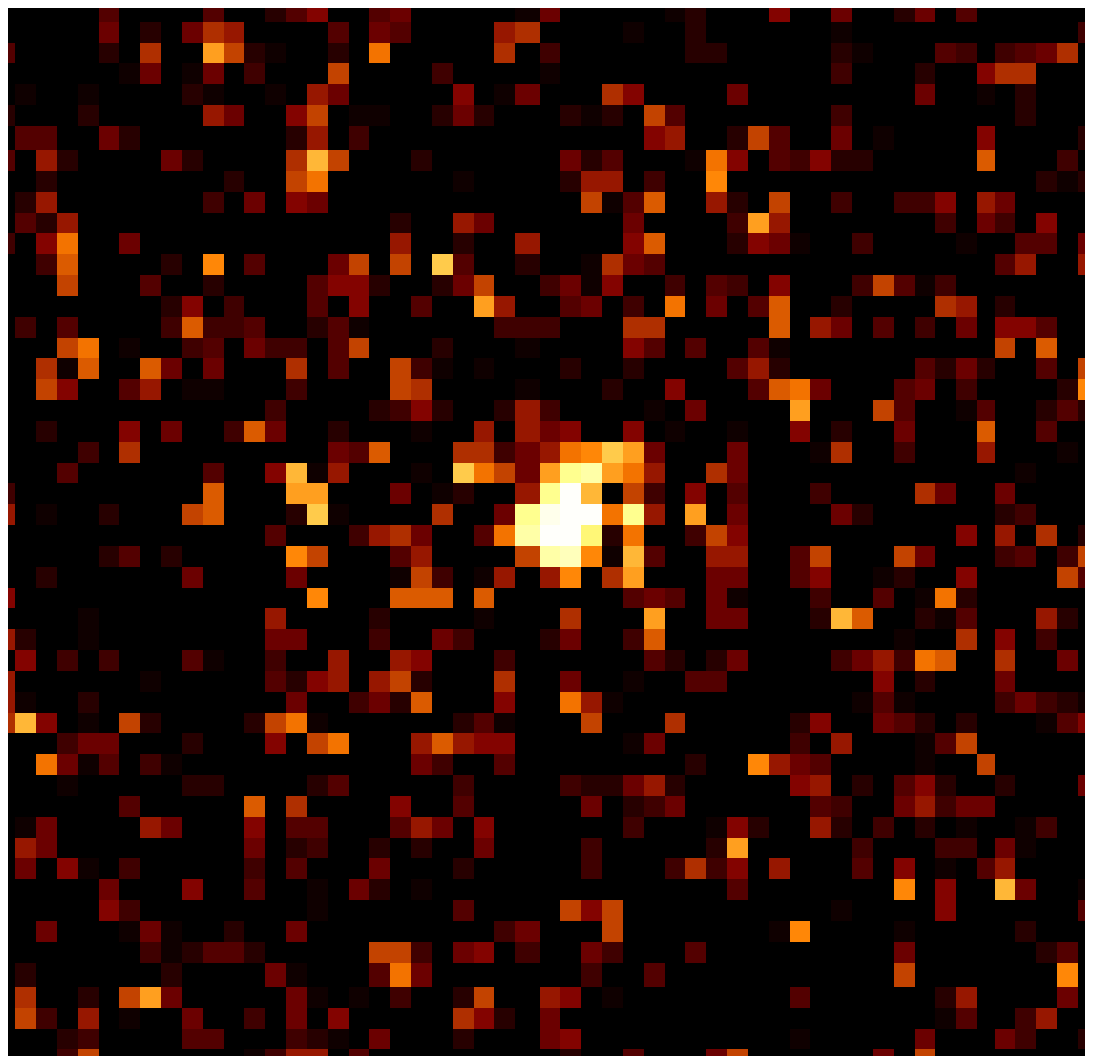, width=5cm}
  \end{center}
\caption{UV image of \object{MCG~-6-30-15}; image approx 25\arcsec \, square, angular resolution 0.5\arcsec }  
\label{ablustin-C2_27_mcg:fig5}
\end{figure}

\object{Markarian~766} (Figure~\ref{ablustin-C2_27_mrk:fig4}; z=0.012929, image approx. 38\arcsec \, square) and \object{MCG~-6-30-15} (Figure~\ref{ablustin-C2_27_mcg:fig5}; z=0.007749, image approx. 25\arcsec \, square), both Narrow Line Seyfert 1s, show far less extended UV emission from their host galaxies than the other objects described here, although \object{Markarian~766} does exhibit UV emission from its spiral arms, about 10\arcsec \, across.

\begin{acknowledgements}

This work is based on observations obtained with XMM-Newton, an ESA science mission with instruments and contributions directly funded by ESA Member States and the USA (NASA).

\end{acknowledgements}


\begin{thebibliography}{}

\bibitem[\protect\astroncite{Mason et~al.}{2001}]{ablustin-C2_27:mas01}
Mason, K.O. et al.\ 2001, A\&A 365, L36-L44

\end{thebibliography}
\end{document}